\documentclass[aps, 10pt, notitlepage, twocolumn, superscriptaddress,nofootinbib,longbibliography]{revtex4-2}

\usepackage{dcolumn}% Align table columns on decimal point
\usepackage{bm}% bold math
\usepackage{ifpdf}
\usepackage{hyperref}
\usepackage{float}
\usepackage{bm}
\usepackage{xcolor,color,graphicx,graphics}
\usepackage[spanish,english]{babel}%, portuguese
\usepackage[latin1]{inputenc}
\usepackage[OT1]{fontenc}
\usepackage{latexsym,amssymb,amsmath,amsfonts}
\usepackage{makeidx}
\usepackage{epsfig,subfigure}
\usepackage{epstopdf}
\usepackage{mathrsfs}
\hypersetup{colorlinks=true, linkcolor=blue, citecolor=green}
\usepackage{enumerate}
\usepackage{xcolor}
 \usepackage{multirow}
\definecolor{red}{rgb}{1,0,0}

\def\+{^\dagger}

\def\<{\leftarrow}
\def\>{\rightarrow}

\def\({\left(}
\def\){\right)}

\def\arcsinh{\mathop{\rm arcsinh}\nolimits}

\newcommand{\bi}{\begin{itemize}} 				\newcommand{\ei}{\end{itemize}}
\newcommand{\benu}{\begin{enumerate}} 		\newcommand{\enu}{\end{enumerate}}
\newcommand{\bd}{\begin{dinglist}{0}}     \newcommand{\ed}{\end{dinglist}}
\newcommand{\bfig}{\begin{figure}[htbp]}  \newcommand{\efig}{\end{figure}}
        			
\newcommand{\bc}{\begin{center}} 				  \newcommand{\ec}{\end{center}}
\newcommand{\be}{\begin{equation}} 				\newcommand{\ee}{\end{equation}}
\newcommand{\bsub}{\begin{subequations}}  \newcommand{\esub}{\end{subequations}}
\newcommand{\ben}{\begin{eqnarray}} 			\newcommand{\een}{\end{eqnarray}}
\newcommand{\ba}[1]{\begin{array}{#1}} 		\newcommand{\ea}{\end{array}}
\newcommand{\bea}{\begin{equation}\begin{array}{rcl}}
\newcommand{\eea}{\end{array}\end{equation}}

\begin{document}

\title{Gravitational lensing and shadows from thin-disks in Loop Quantum Gravity self-dual black holes}

	\author{David J. Pati\~{n}o Pomares} \email{davpatin@ucm.es}
\affiliation{Departamento de F\'isica Te\'orica and IPARCOS,
	Universidad Complutense de Madrid, E-28040 Madrid, Spain}

\author{Diego Rubiera-Garcia} \email{drubiera@ucm.es}
\affiliation{Departamento de F\'isica Te\'orica and IPARCOS,
	Universidad Complutense de Madrid, E-28040 Madrid, Spain}

\date{\today}
\begin{abstract}
We analyze gravitational lensing and their cast images from thin-disks in shadow observations of a family of spherically symmetric black hole solutions previously derived within the framework of Loop Quantum Gravity. Such black holes depend on two parameters (besides the mass of the black hole itself), $P$ and $a_0$, the latter imbuing the configurations with an interior wormhole structure. Using the bounds from the Event Horizon Telescope regarding the shadow's radius of Sgr A$^*$ that constrain the parameter $P \lesssim 0.08(2\sigma)$ (at $a_0=0$),  we study the modifications to weak and strong gravitational lensing induced by these geometries as compared to the Schwarzschild black hole within this range. In particular, we discuss several observables in the strong field regime related to the luminosity decay, the angular separation, and the flux ratio between multiples images of the source. Furthermore, we consider the cast images of these black holes when illuminated by a geometrically and optically thin accretion disk according to several semi-analytic profiles for the disk's emission.

\end{abstract}

\maketitle

\section{Introduction}

One of the most astonishing facts about Einstein's General Relativity (GR) is that the space-time geometry  bends not only trajectories of massive particles but also those of light. This was actually the first test that gave experimental support to GR via Sobral and Eddington's observations of the deflection of light passing by the Sun in 1919's eclipse (see 
 \cite{Crispino:2018yhp} for an historical account). In more modern times, this deflection of light has become a powerful tool for testing the nature of compact bodies in astrophysics \cite{Virbhadra:2002ju,Chen:2009eu,Eiroa:2010wm,Tsukamoto:2016qro,Jusufi:2017mav,Cunha:2018acu,Nascimento:2020ime,Islam:2020xmy,Tsukamoto:2021caq,Kuang:2022xjp,Guo:2022muy}, and finds also many applications within cosmology (see \cite{LensingBook} for a detailed account of such applications). Indeed, since the gravitational lensing effect is greatly exaggerated near compact bodies such as neutron stars and black holes, a variety of methods has been developed in order to address this challenge, starting with the seminal work by Bozza to extract observables of black holes in the strong-field gravitational regime \cite{Bozza:2001xd}. 
 
On the other hand, when the main source of light around a compact enough body (typically a black hole) is provided by the accretion disk surrounding it, the generalization of light deflection of a single ray towards a whole bunch of them leads to new physical observables within the field of  the so-called {\it shadows} \cite{Falcke:1999pj}. The measurement, by the Event Horizon Telescope (EHT) Collaboration, of the image of the supermassive objects at the heart of the M87 \cite{EventHorizonTelescope:2019dse} and Milky Way galaxies \cite{EventHorizonTelescope:2022wkp}, are broadly consistent with the expectation of a rotating black hole surrounded by a magnetized, super-heated plasma, the link between observation and theory provided by the numerical resolution within General Relativistic Magneto-Hydrodynamic (GRMHD) simulations of the accretion flow. In turn, this further bolsters our confidence on the validity of GR to describe these extreme phenomena.

From a theoretical perspective, the theorems on uniqueness, combined with the no-hair conjecture, tell us that the most general asymptotically flat, stationary black hole within GR is given by the Kerr family \cite{Penrose:1969pc,Carter:1971zc}, solely described by mass and angular momentum (augmented to the Kerr-Newman family when an electric charge is included, though such a charge is typically regarded as negligible for any astrophysical application). However, black holes within GR are doomed to hold a space-time singularity inside them, as given by the fact that some sets of geodesics unavoidably become incomplete (for a review and discussion of this topic see \cite{Senovilla:2014gza}). As singularities undermine the predictability of GR, a great deal of activity has been carried out in the literature to supersede GR on its strong-field regime, where supposed quantum-gravity effects should arise to resolve space-time singularities and this way restore the full predictability of the theory. This should be met, in the observational side, with casting specific predictions for astrophysical bodies that could be compared to those of GR ones \cite{Cardoso:2019rvt}, for instance via gravitational lensing, shadows, gravitational waves, and so on.

The main aim of this work is to study the modifications to the geodesic lensing in relation to its applications to shadow images,  of a family of asymptotically flat, spherically symmetric solutions derived in Ref. \cite{Modesto:2008im} and framed within one of the main proponents to quantize gravity: Loop Quantum Gravity (LQG) \cite{Rovelli:1997yv}. This is a non-perturbative quantization of GR which has had several successes within the singularity resolution problem, see e.g. \cite{Bojowald:2001xe,Ashtekar:2003hd}. The modified metric employed here depends on two theory's parameters (besides the mass of the black hole itself), $P$ and $a_0$, playing different roles in the structure of the corresponding solutions. We shall characterize the weak and strong deflection limits of this geometry, and study the evolution of several observables associated to multiple images for values of $P$ within the interval  $P \lesssim 0.08$, resulting from the inferred shadow radius of Sgr A$^*$ \cite{EventHorizonTelescope:2022xqj}, the radio source of the supermassive object at the heart of our Milky Way galaxy, found within $2\sigma$ uncertainty (and assuming $a_0=0$)  \cite{Vagnozzi:2022moj}. Such observables are the Lyapunov exponent of unstable bound geodesics, and the angular separation and flux ratio of multiple images. For the shadow images, we shall generate full images using a geometrically and infinitesimally thin model of the disk, emitting monochromatically in the disk's frame using several semi-analytic profiles, and compute the ratio of intensities between the first and second photon rings of highly-bent light trajectories around LQG black holes, discussing how they relate to theoretical computations based on the Lyapunov exponent above. 

This work is organized as follows.  In Sec. \ref{C:II} we introduce the LQG black hole solutions we are interested in. In Sec. \ref{C:III} we consider the equations of null geodesic motion and particularize them for LQG black holes on the weak and strong field regimes. Some observables associated to such black holes are discussed in Sec. \ref{C:IV}, and full images of thin accretion disk are generated in Sec. \ref{C:V}. We finally depict our conclusion in Sec. \ref{C:VI}.

\section{Loop Quantum Gravity black holes} \label{C:II}

\subsection{Line element}

For the purposes of casting the theoretical framework of this work we shall consider a static, spherically symmetric line element suitably written as
\begin{equation} \label{genmetric}
ds^2 = -A(r)dt^2 + B(r)dr^2 + C(r) d \Omega^2
\end{equation}
where $\{A(r), B(r), C(r) \}$ are general coefficients that depend on $r$ only, while $d\Omega^2=d \theta^2 + \sin^2 \theta d \phi^2$ is the line element on the two-spheres. We point out that while it is always possible to redefine the coordinate $r$ as $B(r)dr^2=dx^2/A(x)$ to get rid of the function $B(r)$ in the above line element in terms of the function $A(r)$, such a change usually typically spoils a simple representation for the function $C(r)$, whose non-trivial behaviour (when present) gives rise to new interesting possibilities departing from canonical black holes.

In \cite{Modesto:2008im} a particular functional form for such coefficients was found, derived from LQG, and given by 
\begin{eqnarray} 
A(r) &=& \frac{(r - r_{+})(r - r_{-})(r + r_{*})^2}{r^4 + {a_{0}}^2} \label{radialcoefficientsone} \\
B(r) &=& \frac{(r^4 + {a_{0}}^2)(r + r_{*})^2}{(r - r_{+})(r - r_{-})r^4} \label{radialcoefficientstwo} \\
C(r) &=& r^2 \left(1+ \frac{{a_{0}}^2}{r^4}\right) \label{radialcoefficientsthree}
\end{eqnarray}
where the parameters characterizing them are given by
\begin{eqnarray}
r_+ &=&  2m \label{eq:rp} \\
r_- &=& 2mP^2\label{eq:rm} \\
r_{*} &=& \sqrt{r_+ r-}=2mP
\end{eqnarray}
The theory thus has two new scales: $a_0$ and $P$. The first scale implements a bouncing behaviour in the radial function $C(r)$, something that it is typically interpreted as the signal of a wormhole structure. The second scale is defined as
%the later related to (what?) as Revisar esto
\begin{equation} \label{parametrop}
P=\frac{\sqrt{1+\gamma^2 \delta^2}-1}{\sqrt{1+\gamma^2 \delta^2}+1}
\end{equation}
where $\gamma$ and $\delta$ are the so-called Barbero-Immirzi parameter and polymer parameters, respectively, names reminiscent on the quantization approach followed in the corresponding implementation of LQG ideas. For the sake of our study, we are only concerned about the value and constraints on $P$ (taking $a_0=0$ to deal with modified black holes rather than wormholes), which therefore shall take as a free parameter in our analysis.

On the other hand, the mass parameter $m$ (which appears as an integration constant of the field equations) can be related to the ADM mass of the system by performing series expansions in the $r \to \infty$ limit as
\begin{eqnarray}
g_{tt} &\approx 1&- \frac{2m}{r}(1-P)^2 + \mathcal{O}\left(\frac{1}{r^2} \right) \\
g_{rr} &\approx 1&+ \frac{2m}{r}(1+P)^2+\mathcal{O}\left(\frac{1}{r^2} \right)
\end{eqnarray}
therefore allowing to make the identification
\begin{equation} \label{masas}
M= m(1+P)^2
\end{equation}
in order for $g_{rr}$ to recover the right Schwarzschild behaviour and thus for  $M$ to correspond to the usual ADM mass of the system.  It is therefore convenient to recast the constants of our metric as
\begin{eqnarray}
r_+ &=&  \frac{2M}{(1+P)^2} \label{eq:rp} \\
r_- &=& \frac{2MP^2}{(1+P)^2} \label{eq:rm} \\
r_{*} &=& \sqrt{r_+ r-}=\frac{2MP}{(1+P)^2}
\end{eqnarray}
so everything gets parameterized in terms of the physical mass determining the actual motion of bodies from the point of view of the asymptotic observer. This will allow us to compare LQG and Schwarzschild black hole quantities on an equal-footing. These are the expressions for the geometrical background we shall use throughout this work. We point out that these configurations are typically dubbed as {\it self-dual} under the so-called T-duality, meaning that under the transformation $r \to a_0/r$, and under suitable reparametrization of its variables, the metric coefficients remain the same \cite{Modesto:2009ve}.

As for the values of the parameter space of $P$ explored here, in 2022 the EHT Collaboration reported that it is possible to derive constraints on the central brightness depression's size (which is not directly observable) of Sgr A$^*$, the radio source at the center of our Milky Way galaxy and consistently interpreted as generated by a supermassive black hole,  via a correlation with the size of the ring of bright radiation surrounding it (which is observable) \cite{EventHorizonTelescope:2022xqj}. While there are some caveats to such a correlation and the calibration procedures used to establish it, this result was  used in \cite{Vagnozzi:2022moj} to set bounds on the values of $P$ (at $a_0=0$) as given by
\begin{eqnarray}
&& 0 \lesssim P \lesssim 0.05 \quad (\text{at} \quad 1\sigma) \\
&& 0 \lesssim P \lesssim 0.08 \quad (\text{at} \quad 2\sigma) \label{eq:con1}
\end{eqnarray}
and for the sake of this work we shall take the second constraint.

\section{Gravitational lensing}  \label{C:III}

Gravitational lensing can be divided into two different regimes, weak and strong, to which we shall apply different techniques.  On the weak-field regime, the trajectory may be distorted slightly, so a distance source will be altered on its apparent location. On the strong-field regime, corresponding to the scenario in which the light ray gets close enough to an unstable bound orbit (the photon sphere for a spherically symmetric metric), it can turn many times around the black hole producing multiple images of the source on the observer's screen. Our analysis here (and the one of Sec. \ref{C:IV}) largely extends the one carried out in Ref. \cite{Sahu:2015dea}. Before going to tackle each regime separately, we go first to derive the equations of motion of null particles.

\subsection{Equations of null geodesic motion}

A particle moving on a null geodesic (in our case a light ray) has a line element given by $ds^2 = 0$.  For a static, spherically symmetric line element of the form (\ref{genmetric}), the fact that the metric coefficients do not depend on $t$ and $\phi$ implies that there are two conserved quantities of motion given by \cite{butterfield2005symmetry}
\begin{equation} \label{conservedquantities}
A(r) \Dot{t} = E, \hspace{0.5 cm} C(r) \Dot{\phi} = L
\end{equation}
interpreted as the energy and angular momentum per unit mass, respectively. Using these conserved quantities, the photon's trajectory equation can be  rewritten as
\begin{equation} \label{lagrangaian}
-A(r) \Dot{t}^2 + B(r) \Dot{r}^2 + C(r) \Dot{\phi}^2 = 0
\end{equation}
This equation can be suitably rewritten as
\begin{equation} \label{fierre}
\frac{d\phi}{dr}= \pm \frac{b}{C(r)}\frac{\sqrt{A(r)B(r)}}{\sqrt{1-b^2\frac{A(r)}{C(r)}}}
\end{equation}
where we have defined $b\equiv L/E$ as the photon's impact parameter. Eq.(\ref{fierre}), together with the expressions (\ref{radialcoefficientsone}), (\ref{radialcoefficientstwo}), and (\ref{radialcoefficientsthree}), provide the deflection angle for every static, spherically symmetric geometry. However, Eq.(\ref{fierre}) can be written in the alternative form 
\begin{equation} \label{fierredos}
\frac{d\phi}{dr}= \pm \frac{1}{\sqrt{\frac{C(r)}{B(r)}\left(\frac{C(r)}{A(r)b^2} - 1\right)}}
\end{equation}
which will make some computations along this work more straightforward.

\subsection{Weak gravitational lensing}

In the weak gravitational lensing regime the photon's source real location is slightly deflected by the compact body producing an apparent location on the observer's screen. In order to integrate the trajectory along the path of the photon from its position to the deflection point and back to the asymptotic observer we assume extremely large distances in both cases. This way, we can integrate Eq.(\ref{fierredos}) from asymptotic infinity to the deflection point $r_0$ twice, i.e.:
\begin{equation} \label{weaklensingintegral}
I_{D} (r_{0}) = 2\int_{r_{0}}^{\infty} \frac{dr}{\sqrt{\frac{C(r)}{B(r)}\left(\frac{C(r)}{A(r)b^2} - 1\right)}}
\end{equation}

To perform the integration for the LQG black holes above, we take the metric functions  (\ref{radialcoefficientsone}), (\ref{radialcoefficientstwo}) and (\ref{radialcoefficientsthree}) and set $a_{0} = 0$, since, as we mentioned above, we are interested in the usual black hole scenario. This way, we obtain the following expression
\begin{equation} \label{weaklensingintegralLQG}
I_{D} (r_{0}) = 2\int_{r_{0}}^{\infty} \frac{dr}{F(r)}
\end{equation}
with the definition
\begin{eqnarray} 
F(r) &=& \sqrt{M(r) - H(r)}\label{coeficientef} \\
M(r) &=& \frac{(r_{0} - r_{-})(r_{0} - r_{+})(r_{0} + r_{*})^2}{r_{0}^{6}} \frac{r^{8}}{(r + r_{*})^4} \label{coeficientem} \\
H(r) &=& \frac{(r - r_{-})(r - r_{+})r^2}{(r + r_{*})^2}\label{coeficienteh}
\end{eqnarray}
We next introduce the dimensionless variable 
\begin{equation} \label{cambiodevariable}
\omega = \frac{r_{0}}{r}
\end{equation}
in terms of which the above expressions turn into 
\begin{eqnarray} 
M(r) &=& \frac{r_{0}^2}{\omega^4} \frac{\left(1 - \frac{r_{+}}{r_{0}}\right)\left(1 - \frac{r_{-}}{r_{0}}\right)\left(1 + \frac{r_{*}}{r_{0}}\right)^{2} }{\left(1 + \frac{\omega r_{*}}{r_{0}}\right)^4} \label{coeficientemnuevo} \\ 
H(r) &=& \frac{r_{0}^2}{\omega^2} \frac{\left(1 - \frac{\omega r_{+}}{r_{0}}\right)\left(1 - \frac{\omega r_{-}}{r_{0}}\right)}{\left(1 + \frac{\omega r_{*}}{r_{0}}\right)^2} \label{coeficientehnuevo}
\end{eqnarray}
Now, taking into account the fact that we are in the weak gravitational lensing regime, in which $r_{0}$ is assumed to be huge, this  transforms (\ref{coeficientef}) into
\begin{eqnarray}
F(\omega) &=& \sqrt{\frac{r_{0}^2}{\omega^4} \left[1 - \omega^2 + \frac{J(\omega)}{r_{0}} + \mathcal{O}\left(\frac{1}{r_{0}^2}\right)\right]} \label{coeficientefmodificado} \\
J(\omega) &=& 2r_{*}(1 - 2 \omega + \omega^{3}) + (r_{+} + r_{-})(\omega^{3} - 1)\label{jota}
\end{eqnarray}
where we have considered all the quadratic terms in $r_0$ appearing in (\ref{coeficientefmodificado}) as negligible, in agreement with the approximation of the weak lensing.

With the approximations and expressions above, we can now integrate  (\ref{weaklensingintegralLQG}), keeping only linear terms. We then find
\begin{equation} \label{integralweakaproximada}
\begin{split}
I_{D} (r_{0}) & \approx \int_{0}^{1} \frac{2 d \omega}{\sqrt{1 - \omega^2}} \left(1 - \frac{J(\omega)}{2r_{0}(1 - \omega^2)}\right) \\
& \approx \pi + \frac{2(r_{+} + r_{-})}{r_{0}}
\end{split}
\end{equation}
which is the expression for weak lensing in LQG black holes. To get the angular deflection suffered by the light ray on its path we must subtract $\pi$ from (\ref{integralweakaproximada}), since it simply tells us that the light ray crossed the entire plane on its path from source to observer.

It is more suitable to rewrite the above expression in terms of the impact parameter $b$. Thus, from the definition of the impact parameter and taking a series expansion, we find
\begin{equation} \label{erreceroconb}
\frac{1}{r_{0}} = \frac{1}{b} - \frac{2r_{*} - r_{+} - r_{-}}{2b^2} + \mathcal{O} \left(\frac{1}{b^3}\right)
\end{equation}
Therefore, the deflection angle (subtracting the $\pi$ factor) in terms of $b$ is given by 
\begin{equation} \label{weakangulardeflection} 
\alpha (r_{0}) \approx \frac{2(r_{+} + r_{-})}{b} = \frac{4M}{b}\frac{1 + P^2}{(1 + P)^2}
\end{equation}
Retaining only linear terms in $P$, then Eq.(\ref{weakangulardeflection}) takes the following form
\begin{equation} \label{weakangulardeflectionaproximada}
\alpha (r_{0}) \approx \frac{4M}{b}(1 - 2P)
\end{equation}
where we see that the corrections introduced by the constant $P$ (assumed to be positive in agreement with the constraint (\ref{eq:con1}))  make the deflection angle smaller than in the Schwarzschild case. Obviously, in the case $P=0$ we recover the well known deflection angle of the Schwarzschild black hole  \cite{Soares:2023uup}. 

\subsection{Strong field limit}

\begin{figure}[t!]
\includegraphics[width=8.4cm,height=5.5cm]{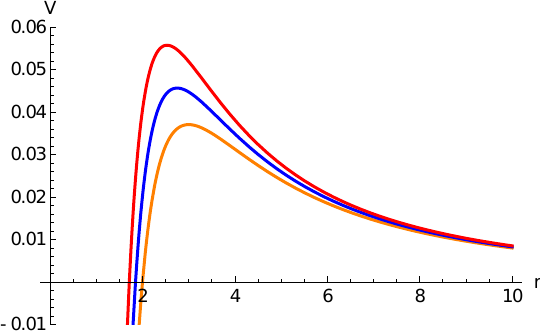}
\caption{The effective potential of null geodesics $V(r)$ in Eq.(\ref{potencialcentral}) for LQG black hole with $\{a_0=0,M = 1\}$ for the parameters $P=0$ (orange, Schwarzschild case), $P=0.04$ (blue) and $P=0.08$ (red).}
\label{fig:potential}
\end{figure}

In the strong deflection limit the photon can turn multiple orbits around the black hole. To characterize this regime we first suitably rewrite the geodesic equation (\ref{lagrangaian}) as (here we re-absorb a factor $L^2$ via a re-definition of the affine parameter) 
\begin{equation} \label{eq:geoeqeff}
AB \dot{r}^2=\frac{1}{b^2}-V(r)
\end{equation}
which is akin to the equation of motion of a one-dimensional particle in the effective potential
\begin{equation} \label{potencialcentral} 
V(r) = \frac{A(r)}{C(r)}
\end{equation}
In Fig. \ref{fig:potential} we depict the effective potential as a function of $r$ for different values of the $P$ parameter as compared to the Schwarzschild solution.

The maximum of the effective potential corresponds to the locus of unstable bound orbits and is known as the {\it photon sphere}. To find it, taking a derivative of the potential (\ref{potencialcentral})  with respect to the radial coordinate and making it equal to zero allows us to obtain the following equation
\begin{equation}
\gamma (r) = \frac{6(r - r_{-})(r - r_{+})(r + r_{*})}{r} 
\label{ecuacioncirculares} 
\end{equation}
where we have introduced the following definitions
\begin{eqnarray}  
\gamma (r) &=& \lambda (r) + \eta (r) \\
\lambda (r) &=& (r - r_{-})(r + r_{*}) + (r - r_{+})(r + r_{*}) \\
\eta (r) &=& 2(r - r_{-})(r - r_{+})
\end{eqnarray}

An exact solution to (\ref{ecuacioncirculares}) takes a extremely complicated form. However, since we are assuming $P$ small, we can neglect quadratic order terms in $P$, so the equation above takes a more friendly form as
\begin{equation} \label{ecuacioncentralsimplificada} 
r^{3} + 10MPr^2 - 3Mr^2 - 10M^2rP = 0
\end{equation}
This is a cubic equation whose solution $r=r_m$ is
\begin{equation} \label{solucionrprimerorden} 
r_{m} = \frac{M}{2}(\sqrt{9 - 20P + 100P^2} - 10P + 3) 
\end{equation}
which can be approximated as
\begin{equation} \label{solucionrprimerordensinraiz} 
r_{m} \approx 3M \left(1 - \frac{20}{9}P\right)
\end{equation}
The first term in this expression corresponds to the usual location of the photon sphere in the Schwarzschild space-time, while LQG black hole corrections in $P$ decrease slightly its value (within the bounds above).

Let us now tackle the calculation of the strong deflection of light. Here we closely follow the approach of Bozza \cite{Bozza:2010xqn} (see also \cite{Tsukamoto:2016jzh}), from which the light deflection angle can be obtained in the limit where null geodesics approach the photon sphere
(\ref{solucionrprimerordensinraiz}) using the formula
\begin{equation} \label{alfastrong}
\alpha (b) = - \bar{a} \log{\left(\frac{b}{b_{c}} - 1\right)} +  \bar{b}  + \mathcal{O}((b - b_{c}) \log{(b - b_{c})})
\end{equation}
where we have introduced the coefficients (here the subindex $m$ means evaluation at the photon sphere location)
\begin{eqnarray} \label{abarra}
\bar{a} &=& \sqrt{\frac{2 B_{m} A_{m}}{C_{m}^{\prime \prime}A_{m} - C_{m}A_{m}^{\prime \prime}}} \\
 \label{bbarra}
\bar{b} &=& \bar{a} \log{\left[r_{m}^2 \left(\frac{C_{m}^{\prime \prime}}{C_{m}} - \frac{A_{m}^{\prime \prime}}{A_{m}}\right)\right]} + I_{R}(r_{m}) - \pi \label{eq:bparam} \\ \label{bcritica}
b_{c} &=& \lim_{r_{0} \to r_{m}} \sqrt{\frac{C(r_{0})}{A(r_{0})}}
\end{eqnarray}

The last coefficient in these formulae is known as the critical impact parameter, and corresponds to the impact parameter a photon needs to have in order to asymptote to the bound orbit (i.e. to the photon sphere). Therefore, photons equaling its value have formally an infinite deflection angle in Eq.(\ref{fierredos}).

On the other hand, Eq.(\ref{eq:bparam}) contains the regular part of the deflection angle integral. Indeed, the latter is incomplete, but can be regularized as follows. One writes  
\begin{equation} \label{integralregular}
I_{R}(r_{m}) \equiv \int_{0}^{1} f_{R}(z, r_{m}) 
\end{equation}
where the regular part of the integral comes from subtracting the divergent contribution to the full integral, that is
\begin{equation} \label{integrandoregular}
f_{R}(z, r_{m}) \equiv f(z, r_{m}) - f_{D}(z, r_{m})
\end{equation}
These two pieces contributing to the total integral can be isolated as 
\begin{eqnarray}  
f(z, r_{m}) &=& \frac{2 r_{m}}{\sqrt{G(z,r_{m})}} \label{integralentera} \\
G(z,r_{m}) &=& \left(\frac{C_{m}^2}{A_{m} B_{m} b^2} - \frac{C_{m}}{B_{m}}\right)(1 - z)^4 \label{funciongmayuscula}
\end{eqnarray}
for the regular piece and 
\begin{eqnarray}  
f_{D}(z, r_{m}) &=& \frac{2 r_{m}}{\sqrt{C_{1} (r_{m})z + C_{2}(r_{m})z^2}} \label{funciondivergente} \\
C_{1}(r_{m}) &=& \frac{C_{m} D_{m} r_{m}}{B_{m}} \label{coeficientecuno} \\
C_{2}(r_{m}) &=& \frac{C_{m} r_{m}}{B_{m}}\left[K_{m} + \frac{r_{m}}{2}\left(\frac{C_{m}^{\prime \prime}}{C_{m}} - \frac{A_{m}^{\prime \prime}}{A_{m}}\right)\right]  \label{coeficientecdos} \\
K_{m} &=& D_{m}\left[(D_{m} - \frac{B_{m}^{\prime}}{B_{m}})r_{m} - 3\right] \label{cademe} \\
D_{m} &=& \frac{C_{m}^{\prime}}{C_{m}} - \frac{A_{m}^{\prime}}{A_{m}}
\end{eqnarray}
for the divergent one. We recall that these expressions can be particularized to the present case just by substituting the corresponding values of the metric coefficients of LQG black holes discussed in the previous section.  Setting $M = 1$ 
for simplicity, in Table \ref{tablacoeficientesbarra} we depict the values of strong-deflection coefficients (\ref{abarra}) and (\ref{bbarra}) for several values of the $P$ parameter. It is seen that the value of $\bar{a}$ is slightly modified outwards as $P$ is increased, while those modifications on $\bar{b}$ are much more noticeable. 

Our main  interest here is to figure out how the specific parameters of the LQG metric yield deviations with respect to the Schwarzschild behaviour. First, in order to evaluate the function $G(z)$ appearing in the regular part of the integrand of the deflection angle, Eq.(\ref{funciongmayuscula}), we follow  the same strategy as before by considering only linear terms in $P$, which allows us carry out an expansion for the $G(z)$ function in Eq.(\ref{funciongmayuscula}) as
\begin{equation} \label{integrandoaprimerorden}
\begin{split}
G (z) & = -3M^2(-3 z^2 + 2z^3) - \\
& - \frac{8}{3}M^2(18z^2 - 17z^3 + 3z^4)P  
\end{split}
\end{equation}
which smoothly recovers the classical solution when $P = 0$. 
We also need to calculate (\ref{bcritica}) since it appears in  (\ref{alfastrong}). Considering again linear terms in $P$ we find
\begin{equation} \label{bcriticaconp}
b_{c} = \sqrt{3} M(3 - 8P) + \mathcal{O} (P^2)
\end{equation}
which implies the impact parameter factor decreases with increasing $P$ as compared to the Schwarzschild solution. 

\begin{table}[t!]
\centering
\begin{tabular}{|c|c|c|c|c|c|c|}
\hline
$P$ & 0 (GR) & 0.01 & 0.02 & 0.05 & 0.08\\ 
\hline
$\bar{a}$ & 1 & 1.004 & 1.009 & 1.023 & 1.037\\ 
\hline
$\bar{b}$ & -0.400 & -0.492 & -0.808 & -1.076 & -1.496 \\ 
\hline
\end{tabular}
\caption{Coefficients $\bar{a}$ and $\bar{b}$ for increasing values of P.}
\label{tablacoeficientesbarra}
\end{table}

\begin{figure}[t!]
  \begin{center}
    \begin{subfigure}
    	[$P = 0.02$]{
        \includegraphics[width=0.5\textwidth, height=0.20\textheight]{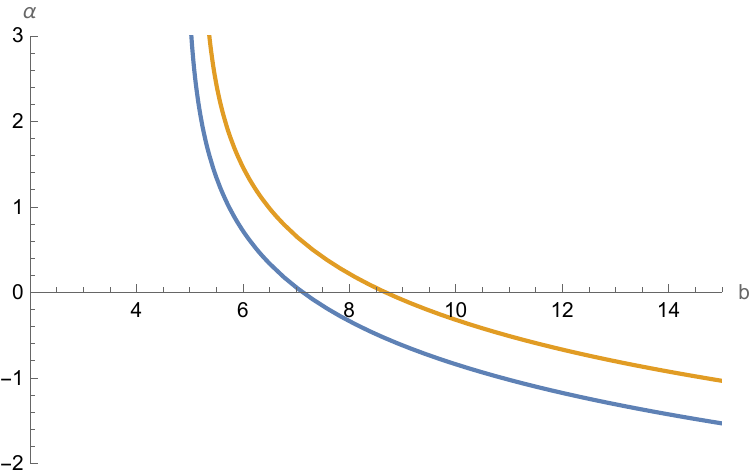}
        \label{graficastrongfieldlimitcinco}}	
    \end{subfigure}
	\begin{subfigure}
		[$P = 0.05$]{
        \includegraphics[width=0.5\textwidth, height=0.20\textheight]{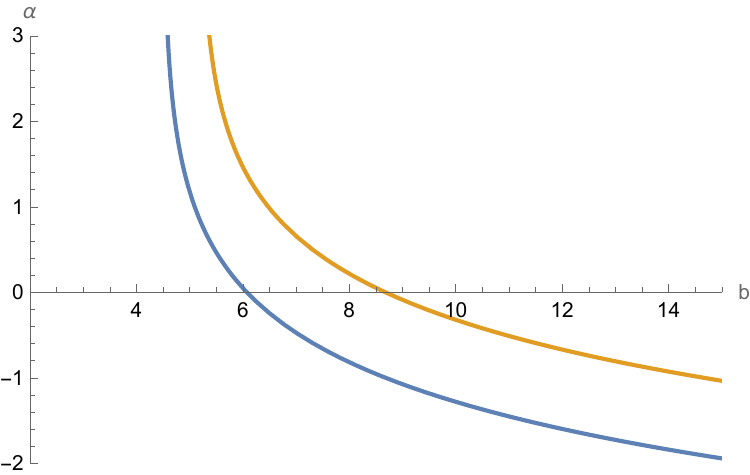}
        \label{graficastrongfieldlimitcien}}
    \end{subfigure}
    \caption{Representation of the light deflection angle as a function of the impact parameter, $\alpha (b)$, in the strong field limit (blue) and in the Schwarzschild case (yellow) for the  values of $P=0.02$ (top) and $P=0.04$ (bottom).}
    \label{graficasstrongfieldlimit}
  \end{center}    
\end{figure}

With the expressions above, in Fig. \ref{graficasstrongfieldlimit} we depict the result of the numerical integration of the deflection angle for values of  $P=0.02$ and $P=0.05$, respectively. We find that the deflection angle increases with growing $P$ as compared to its Schwarzschild counterpart as its critical parameter (which is also reduced) is approached. We can thus interpret this effect as saying that LQG black holes bend space-time slightly stronger than their Schwarzschild counterpart.

\section{Lensing observables} \label{C:IV}

Once we have studied the gravitational lensing in these two regimes, in this section we study some observables as a way to connect physical measurements with predictions of the theory.

\subsection{Lyapunov exponent on nearly-bound orbits}

We first consider the trajectories of photons that pass close by the photon sphere $r=r_m$. In such a case, from (\ref{potencialcentral}), assuming an initial location $r=r_m + \delta r$, where $\delta r_0 \ll \delta r_0$, taking a  Taylor series expansion we find
\begin{equation} \label{expansion}
V(r) \approx V(r_{m}) + \frac{1}{2} V''(r_{m}) (\delta  r)^2 + \ldots
\end{equation}
where the first derivative of the potential vanishes due to the imposition of the photon sphere condition. Now,  Replacing (\ref{expansion}) in (\ref{eq:geoeqeff}) we find
\begin{equation} \label{manipulaciondos}
\left(\frac{d \delta r}{d \phi}\right)^2 = - \frac{1}{2} \frac{C^2}{AB} V''(r_{m}) (\delta r)^2
\end{equation}
where one identifies
\begin{equation} \label{doblederivadapotencial}
V''(r_{m}) = \frac{A''_{m}C_{m} - A_{m}C''_{m}}{C^2_{m}}
\end{equation}
We can thus write the above equation as 
\begin{eqnarray}  
\pi \frac{d \delta r}{d \phi} &=& \gamma \delta r \label{ecuacionlyapunov} \\
|\gamma| &=& \pi \sqrt{\frac{1}{2} \frac{AC'' - A''C}{AB}} \label{exponentelyapunov}
\end{eqnarray}
In this expression $\gamma$ is known as the {\it Lyapunov exponent}, which characterizes the growth of the radial distance of the original orbit after a deflection angle $\pi$ has elapsed, i.e., after the photon has turned a number $n \equiv \phi/\pi$ of half-orbits in its path around the black hole before reaching the asymptotic observer. The above equation can be integrated and written in terms of the number of half-orbits $n$ as
\begin{equation} \label{lyapunovresuelto}
\delta r_{n} = e^{\gamma n} \delta r_0
\end{equation}
where $\delta r_0$ is an integration constant setting the original location of the photon. This means that after a number of half-turns $n$ the photon moves away a distance $e^{\gamma n}$ from its original location, and thus the Lyapunov exponent can be understood as a measure of how strong the space-time is bent near the photon sphere. For the LQG metric considered in this work, the Lyapunov exponent reads, at linear order, as
\begin{equation} \label{exponentelyapunovnuestro}
\gamma = \pi \left(1- \frac{4 }{9}P \right) + \mathcal{O}(P^2)
\end{equation}
From this expression we see that larger values of $P$ reduce (\ref{exponentelyapunovnuestro}), which means that radial perturbations grow more slowly in LQG black holes than in Schwarzschild solution. In Table \ref{tablaLya} we depict the Lyapunov index for several values of $P$ within the observational constraints, showing a very mild deviation as compared to the Schwarzschild value $\gamma_S=\pi \approx 3.14$.

\begin{table}[t!]
\centering
\begin{tabular}{|c|c|c|c|c|c|c|}
\hline
$P$ & 0 (Sch) & 0.01 & 0.02 & 0.05 & 0.08\\ 
\hline
$\gamma$ & 3.14 & 3.13 & 3.11 & 3.07 & 3.03 \\ 

\hline
\end{tabular}
\caption{Lyapunov exponent for increasing values of $P$. Here $P=0$ corresponds to the well known value of the Schwarzschild case.}
\label{tablaLya}
\end{table}

\subsection{Multiple images}

\begin{figure}[t!]
\centering
\includegraphics[width=0.45\textwidth, height=0.35 \textheight]{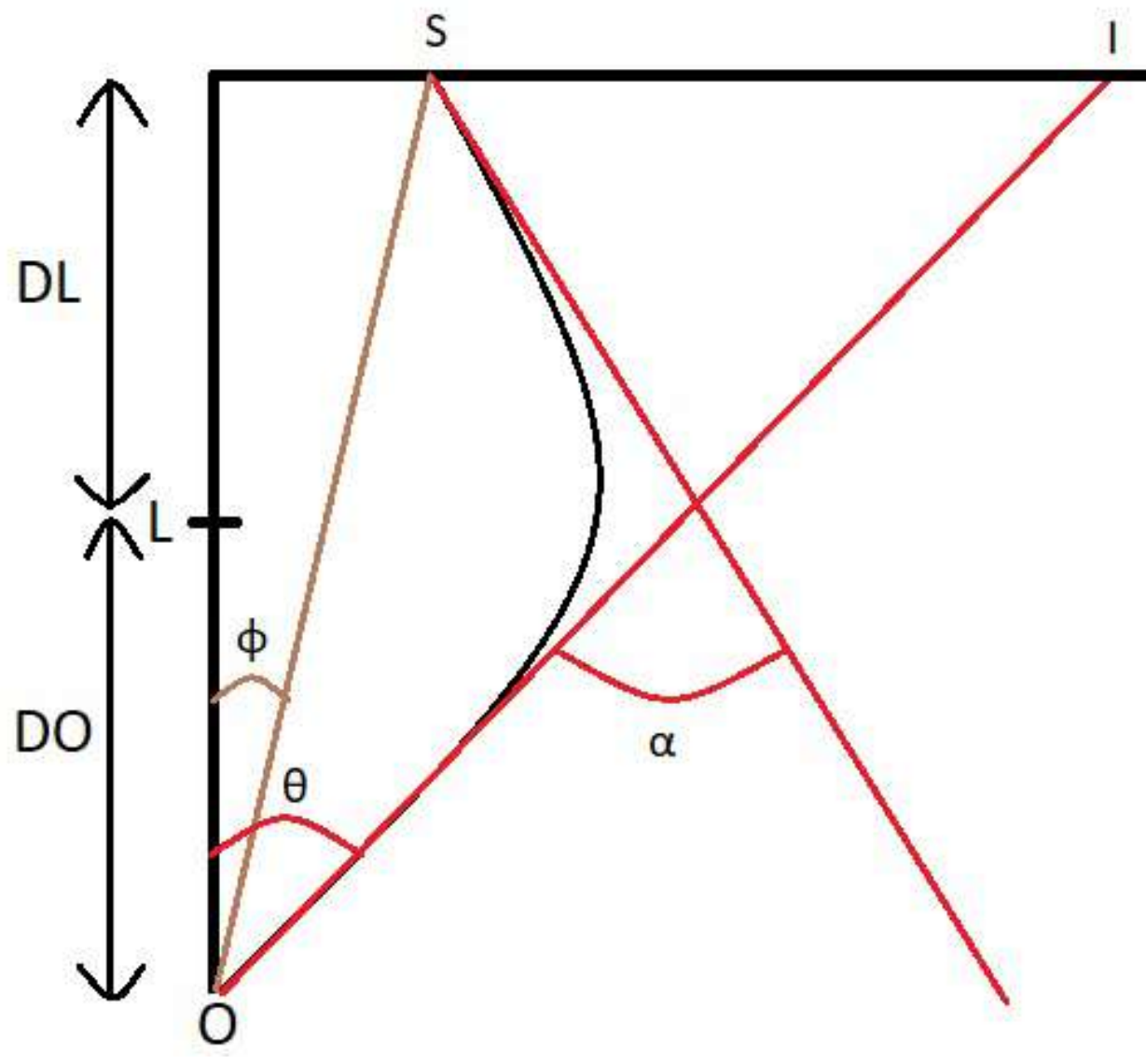}
\caption{Schematic deflection of light. $\theta$ and $\phi$ are the angles the observer measures with respect to the source and the image, respectively. DO and DL are the distances from the observer to the lens, and from the lens to the source, respectively. L marks the spot where the black hole is placed.} 
\label{esquemaobservables}
\end{figure}

Let us now consider another set of observables related to the multiples images created by light trajectories turning $n$ half-times around the black hole. A glance at Fig. \ref{esquemaobservables} tells us about the setup we are interested in: a source $S$ emits a light ray at an angle $\phi$ which, after the bending by the black hole, it is apparently located at $I$ with an angle $\theta$. Basic trigonometric operations allow to relate both angles via the equation
\begin{equation} \label{tangentesobservables}
\tan{\phi} = \tan{\theta} - \frac{DL}{DO} [\tan{\theta} + \tan({\alpha - \theta})]
\end{equation}
Assuming small angles of both source and observer, the above equation becomes
\begin{equation} \label{angulosobservables}
\phi = \theta - \frac{DL}{DO} \Delta \alpha_{n}
\end{equation}
where $\Delta \alpha_{n}=\alpha - 2n \pi$ is the final deflection angle around the black hole after $n$ half-turns. On the other hand, from the definition of the impact parameter and from Fig. \ref{esquemaobservables} we get the expression of the impact parameter as
\begin{equation} \label{impacparameterwithobservables}
b \approx \theta DO
\end{equation}

We are interested here in studying how the angle $\theta$ changes as the number of half-orbits $n$ increases, i.e., as we get closer to the photon sphere. We thus want to expand the values of $\theta$ for which the angular deflection becomes an entire multiple of loops. Therefore, we expand $\Delta \alpha_{n}$ around $\theta = \theta^{0}_{n}$, where $\alpha (\theta^{0}_{n}) = 2n \pi$ and $n$ $\in \mathbb{Z}^{+}$ as
\begin{equation} \label{expansiondeltaalfan}
\Delta \alpha_{n} = \left.\frac{\partial \alpha}{\partial \theta}\right|_{\theta = \theta^{0}_{n}}(\theta - \theta^{0}_{n})
\end{equation}
Now, we substitute (\ref{alfastrong}) and (\ref{impacparameterwithobservables}) into (\ref{expansiondeltaalfan}) and determine $\theta^{0}_{n}$ as
\begin{eqnarray}  
\theta^{0}_{n} &=& \frac{b_{c}}{DO} (1 + e_{n}) \label{thetaceron} \\
e_{n} &=& e^{\frac{\bar{b} - 2n \pi}{\bar{a}}} \label{en} \\
\Delta \alpha_{n} &=& - \frac{\bar{a} DO}{b_{c} e_{n}} (\theta - \theta^{0}_{n}) \label{deltaalfandefinitivo}
\end{eqnarray}
Finally, the critical impact parameter corresponds to those photons that will formally turn an infinite number of times, which makes (\ref{impacparameterwithobservables}) to read in this case as
\begin{equation} \label{bcriticaythetainfinito}
b_{c} = DO  \theta_{\infty}
\end{equation}

We got all the expressions we needed. Let us suppose that all the images obtained for two or more turns after the first one are so close to each other that cannot be distinguished \cite{Bozza:2002zj} (which is actually very close to the physical situation). We can nonetheless compute the angular separation between the angle measured after one turn and all the others combined and approximated by $\theta_{\infty}$. Thus, combining (\ref{thetaceron}), (\ref{en}) and the expression for $b$ leads to the formula
\begin{equation} \label{angularseparation}
s \equiv \theta_{1} - \theta_{\infty} = \theta_{\infty} e^{\frac{\bar{b} - 2\pi}{\bar{a}}}
\end{equation}
The interest on this formula stems from the fact that using it, from observations of these images, one could compute the parameters $\bar{a}$ and $\bar{b}$, thus providing a link of observables with parameters of interest of modified black hole metric, as the $P$ parameter considered in LQG black holes.

In Table \ref{tablaseparacionesangulares} we depict the evolution of the angular separation with $P$ using the values of $\bar{a}$ and $\bar{b}$ previously found in Table \ref{tablacoeficientesbarra}. We observe a lowering of the angular separation with increasing $P$, so higher-order loops beyond the first one would appear closer in the images from the first ring as compared to Schwarzschild black hole. This is in agreement with our analysis of the Lyapunov exponent, which revealed that increasing values of $P$ made the radial perturbations to grow more slowly than in the Schwarzschild case. 

%If the photon describes at least two loops, it will be more deviated from the one loop case for bigger values of $P$ as a consequence of travelling one more time around the black hole. 

\begin{table}[t!]
\centering
\begin{tabular}{|c|c|c|c|c|c|c|}
\hline
$P$ & 0 (Sch) & 0.01 & 0.02 & 0.05 & 0.08\\ 
\hline
$s$ & 0.0065 & 0.0060 & 0.0044 & 0.0034 & 0.0023 \\ 
\hline
\end{tabular}
\caption{Angular separation of LQG black holes for increasing values of $P$ taking $M = DO =1$. Here $P=0$ corresponds to the Schwarzschild case.}
\label{tablaseparacionesangulares}
\end{table}

Another relevant observable of multiple images is the flux ratio. A compact body is able to modify the trajectory of light, but not the surface brightness it carries with it. Nonetheless, the gravitational field affects the solid angle of the source distorting it. The magnification of this geometric element is defined as follows \cite{Bozza:2001xd}
\begin{equation} \label{definiciondemagnification}
\mu_{n} \equiv  \frac{d \Omega_{observed}}{d \Omega_{source}}
\end{equation}
and remembering the lens equation (\ref{angulosobservables}), it can be expressed as
\begin{equation} \label{definiciondemagnificationlarga}
\mu_{n} =  \frac{1}{|\left.det J\right|_{\theta^{0}_{n}}} = \frac{\theta^{0}_{n}}{\phi \frac{\partial \phi}{\partial \theta}\Big|_{\theta^{0}_{n}}}
\end{equation}
Obviously, the magnification tends to infinity if $\phi \rightarrow 0$, and that occurs when the source is perfectly aligned with the lens. From it one defines the flux ratio as \cite{Molla:2023yxn}
\begin{equation} \label{fluxratio}
r \equiv  \frac{\mu_{1}}{\sum_{n = 2}^{\infty} \mu_{n}} \approx \frac{5 \pi}{\bar{a} \log{10}}
\end{equation}

In Table \ref{tablafluxratio} we compute the flux ratio for several values of the $P$ parameter of LQG black holes. We see that the flux ratio becomes smaller with growing $P$, which should not come as a surprise since Eq.(\ref{definiciondemagnification}) indicates that a bigger solid angle observed implies a larger magnification. 

\begin{table}[t!]
\centering
\begin{tabular}{|c|c|c|c|c|c|c|}
\hline
$P$ & 0 (Sch) & 0.01 & 0.02 & 0.05 & 0.08\\ 
\hline
$r$ & 918 & 890 & 863 & 786 & 715 \\ 
\hline
\end{tabular}
\caption{Flux ratio of LQG black holes for increasing values of $P$ taking $M = 1$.  Here $P=0$ corresponds to the Schwarzschild case.}
\label{tablafluxratio}
\end{table}

\section{Imaging LQG black holes}  \label{C:V}

In this section we consider the full images generated by LQG black holes when illuminated by an optically and geometrically thin accretion disk. Such images are generated by the collection of all light rays bent in the gravitational field of the black hole, and thus it is anchored in the study of gravitational lensing carried out in the previous sections.

In the observer's plane image, light rays  above critical impact parameter asymptote to the photon sphere when traced backwards. Therefore, when one performs a ray-tracing analysis of the geometry by which Eq.(\ref{fierre}) is integrated backwards (i.e. with the $-$ sign) one of these two situations will happen: i) if $b>b_c$ a turning point is reached as given by the zeros in the denominator of such an equation and then the integration is further carried out with the $+$ sign until asymptotic infinity is reached again, or ii) if $b<b_c$ the ray intersects the event horizon and the integration is stopped there. The apparent curve in the observer's plane image splitting both kinds of trajectories with $b=b_c$ is dubbed as the critical curve, and corresponds to the projection of the photon sphere there. This allows to define a  {\it mathematical shadow} as given by the sharp decrease in the brightness received in the observer's screen for values of $b<b_c$. However, this does not exactly coincide (except in rather unrealistic scenarios of fully homogeneous illumination) with the image shadow, as given by the features of the central brightness depression caused by the actual properties of the disk and, more particularly, on its geometrical and optical properties. Indeed, only in spherically symmetric geometries of the disk will the outer edge of the central brightness depression coincide with the critical curve (i.e. with the mathematical shadow) \cite{Falcke:1999pj}. For thick disks as long as they are not completely spherical \cite{Vincent:2022fwj}, there will be emission from inside the photon sphere outwards and some light rays will be able to reach the asymptotic observer, producing a neat reduction in the size of the central brightness depression as compared to that of the mathematical shadow, up to a minimum {\it inner shadow} which is solely determined by the properties of the background geometry \cite{Chael:2021rjo}. 

\begin{figure}[t!]
\includegraphics[width=8.4cm,height=5.5cm]{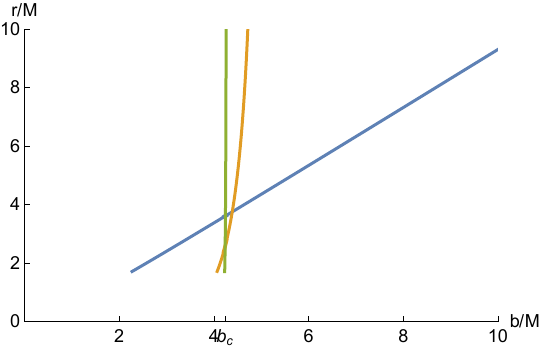}
\caption{The transfer function for LQG black holes taking $P=0.08$ for the direct emission (blue), and the first (orange) and second (green) photon ring emissions.}
\label{fig:tf}
\end{figure}

In order to generate images of LQG black holes in our setting we consider a infinitesimally thin disk with zero opacity. Furthermore, we assume a monochromatic flux in the frame of the disk. Under these conditions, the effect of gravitational redshift between source (disk) and observer can be accounted for using Liouville's theorem, which implies the conservation of the flux in both frames, i.e.,
\begin{equation}
\frac{I_{\nu_o}}{\nu_o^3}=\frac{I_{\nu_e}}{\nu_e^3}
\end{equation}
where $\nu_o$ and $\nu_e$ refer to the frequencies in the observer's and emitter's frames, respectively. Defining the redshift factor $g=\nu_0/\nu_e$, the above equation tells us that $I_{\nu_o}=g^3 I_{\nu_e}$.  Now, integrating over the full spectrum of frequencies in the observer's frame we get 
\begin{equation}
    I_0= \int d\nu_o I_{\nu_o} = \int d\nu_e g^4 I_{\nu_e}=g^4 I(r)
\end{equation}
where we have used the fact that $I_e= \int d\nu_e I_{\nu_e} \equiv I(r)$ given the monochromatic assumption on that frame. On the other hand, using Eq.(\ref{genmetric}) we find that $g=\sqrt{A(r)/A(r_{\infty})}$, where we set the factor $A(r_{\infty}) \to 1$ given the asymptotically flat character of the space-time. This way the observed intensity reads as
\begin{equation}
    I_o(r)=A^2(r) I(r)
\end{equation}
On the other hand we must take into account the fact that photons travelling near the photon sphere are capable to turn more than one half-times around the black hole, this way contributing to the total observed luminosity. Therefore,  the above expression must be corrected to take into account this fact, so we can write the total luminosity received on the observer's screen as 
\begin{equation}
I_o(r)=\sum_{n=0}^{\infty} A^2(r) I(r)
\end{equation}
where now $n$ is understood as counting the number of times the photon's trajectory cuts the equatorial plane (i.e. the accretion disk) beyond the direct emission of the disk (which would correspond to $n=0$). However, as discussed with the Lyapunov exponents in the previous section, the contribution to the total luminosity of successive half-orbits is exponentially suppressed, so after $n=2$ all contributions are completely washed out. This way, for the sake of the generation of our images we shall only take into account the direct ($n=0$) and $n=1,2$ photon ring emissions. To see this, in Fig. \ref{fig:tf} we depict the transfer function (namely, the location of the disk each photon crossed the disk versus the corresponding impact parameter) of LQG black holes for $P=0.08$ for the $n=0, 1, 2$ contributions. Given the fact that the slope of the corresponding curve is associated to the degree of demagnification in the images, it is neatly seen how the direct emission (blue) will dominate the optical appearance as compared to the first (orange) and second (green) photon ring emissions.

\begin{figure}[t!]
\includegraphics[width=8.4cm,height=5.5cm]{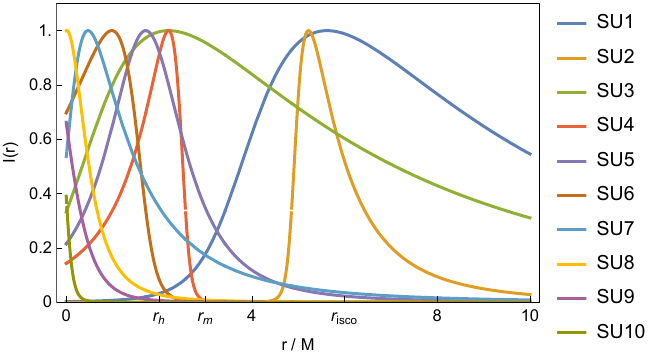}
\caption{The ten SU profiles employed in this work, corresponding to different choices (see Ref.\cite{DeMartino:2023ovj}) of the parameters $\{\gamma,\mu,\sigma\}$ in Eq.(\ref{eq:SU}) for the LQG black holes with $P=0.08$. In this plot $r_h,r_m,r_{ISCO}$ denote the locations of the event horizon, photon sphere, and innermost stable circular radius, respectively. All intensities are normalized to their maximum value.}
\label{fig:Pro}
\end{figure}

For the emission model we employ a technique frequently used in the context of black hole imaging as given by semi-analytic models depending on a finite number of parameters allowing to mimic complex simulations of the accretion flow in GRMHD implementations. Specifically we focus on the family of models introduced by Gralla, Lupsasca and Marrone in \cite{Gralla:2020srx}. These are defined by a suitable adaptation of Johnson's Standard Unbound (hereafter dubbed as SU) distribution, reading explicitly as
\begin{equation} \label{eq:SU}
I(r;\gamma,\mu,\sigma)=\frac{\text{exp}[-\frac{1}{2}(\gamma+\arcsinh (\frac{r-\mu}{\sigma})^2)]}{\sqrt{(r-\mu)^2+\sigma^2}}
\end{equation}
where $\{\gamma,\mu,\sigma\}$ are freely-adjustable parameters, which control different aspects of the emission profile. For the sake of our analysis we shall take ten profiles corresponding to different choices of such parameters (which are those depicted in Table I of Ref. \cite{DeMartino:2023ovj}), and which captures different features of the disk. Such profiles are depicted in Fig. \ref{fig:Pro} for LQG black holes with $P=0.08$, combining different location of the peaks and strengths of the decay with distance, so as to study different scenarios of the emission. In particular, we organize such profiles according to the location of the peak of the emission from the outermost (SU1) to the innermost (SU10). Obviously, only the part of these profiles outside the event horizon $r_h$ is relevant for the generation of images. Besides the peak's location, the decay with the distance will have something to say about the global images.

\begin{figure*}[t!]
\includegraphics[width=4cm,height=3cm]{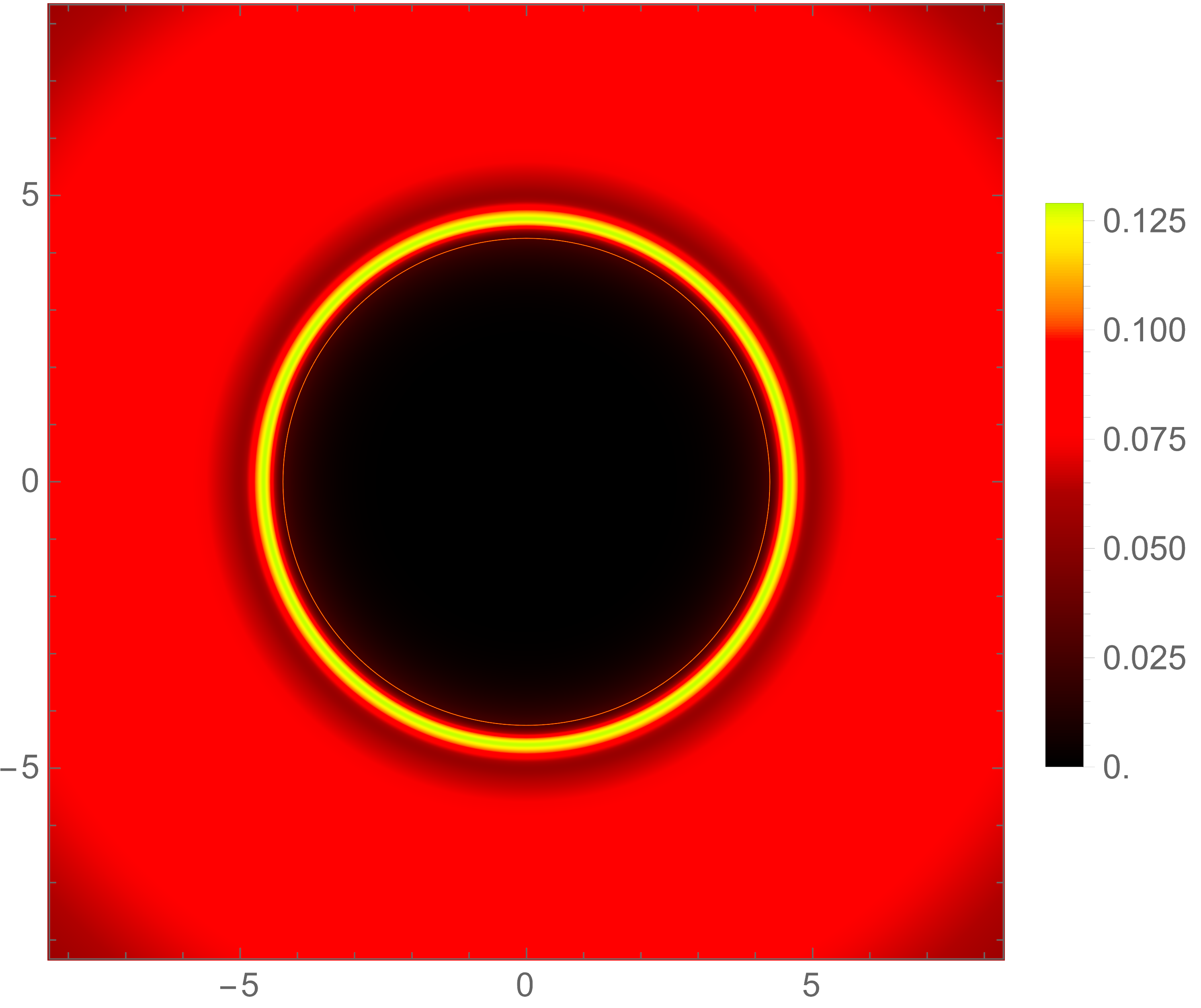}
\includegraphics[width=4cm,height=3cm]{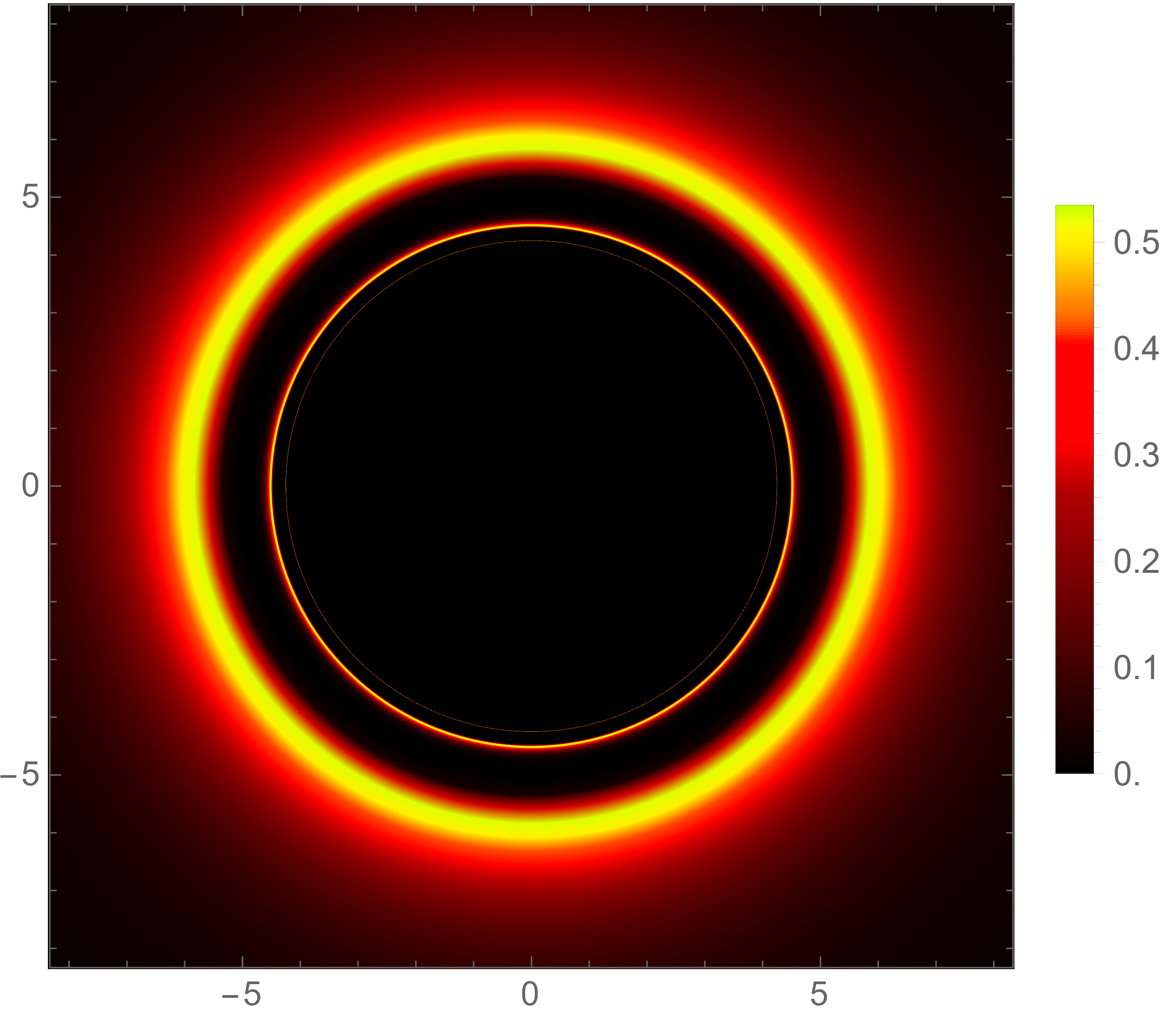}
\includegraphics[width=4cm,height=3cm]{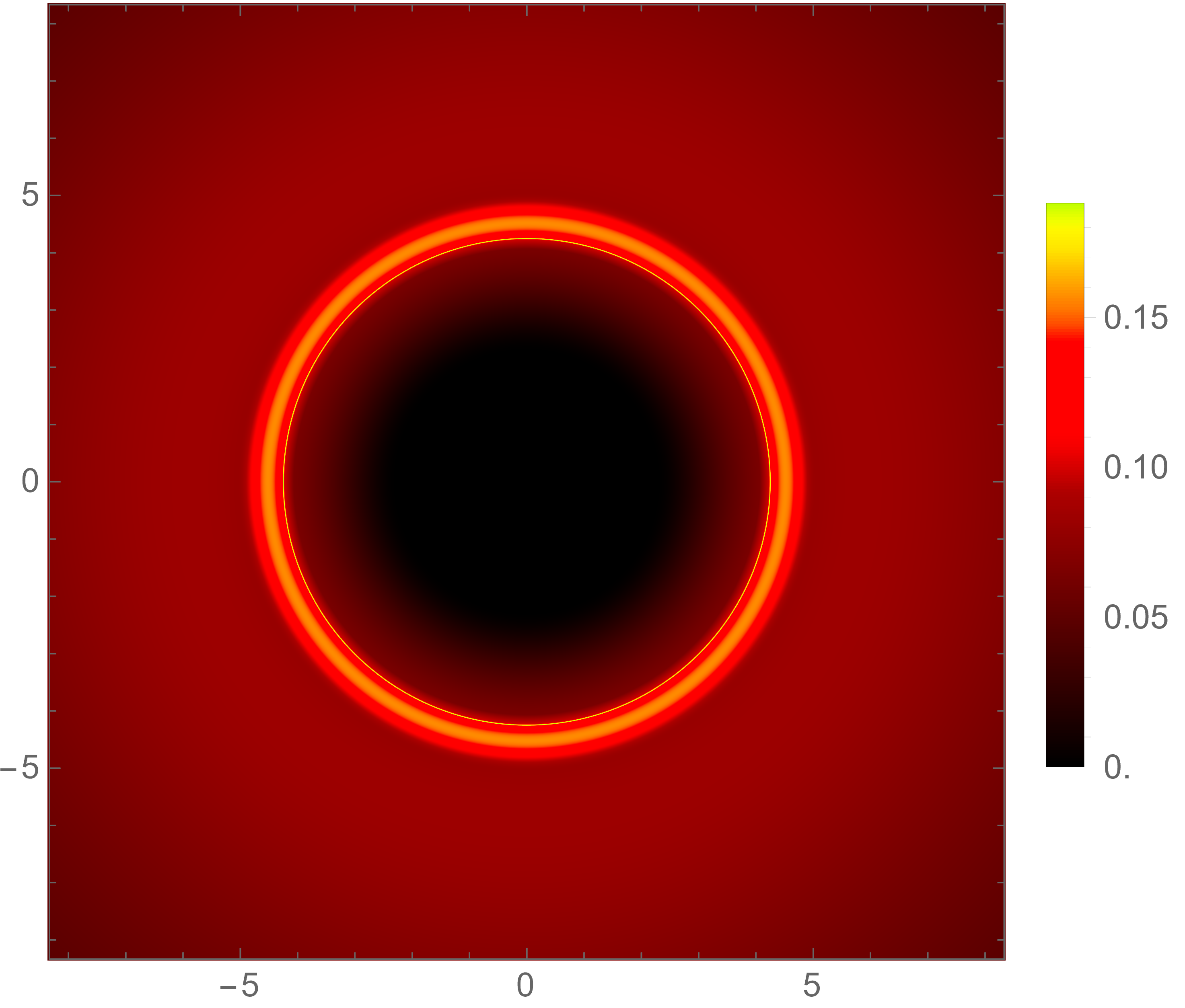}
\includegraphics[width=4cm,height=3cm]{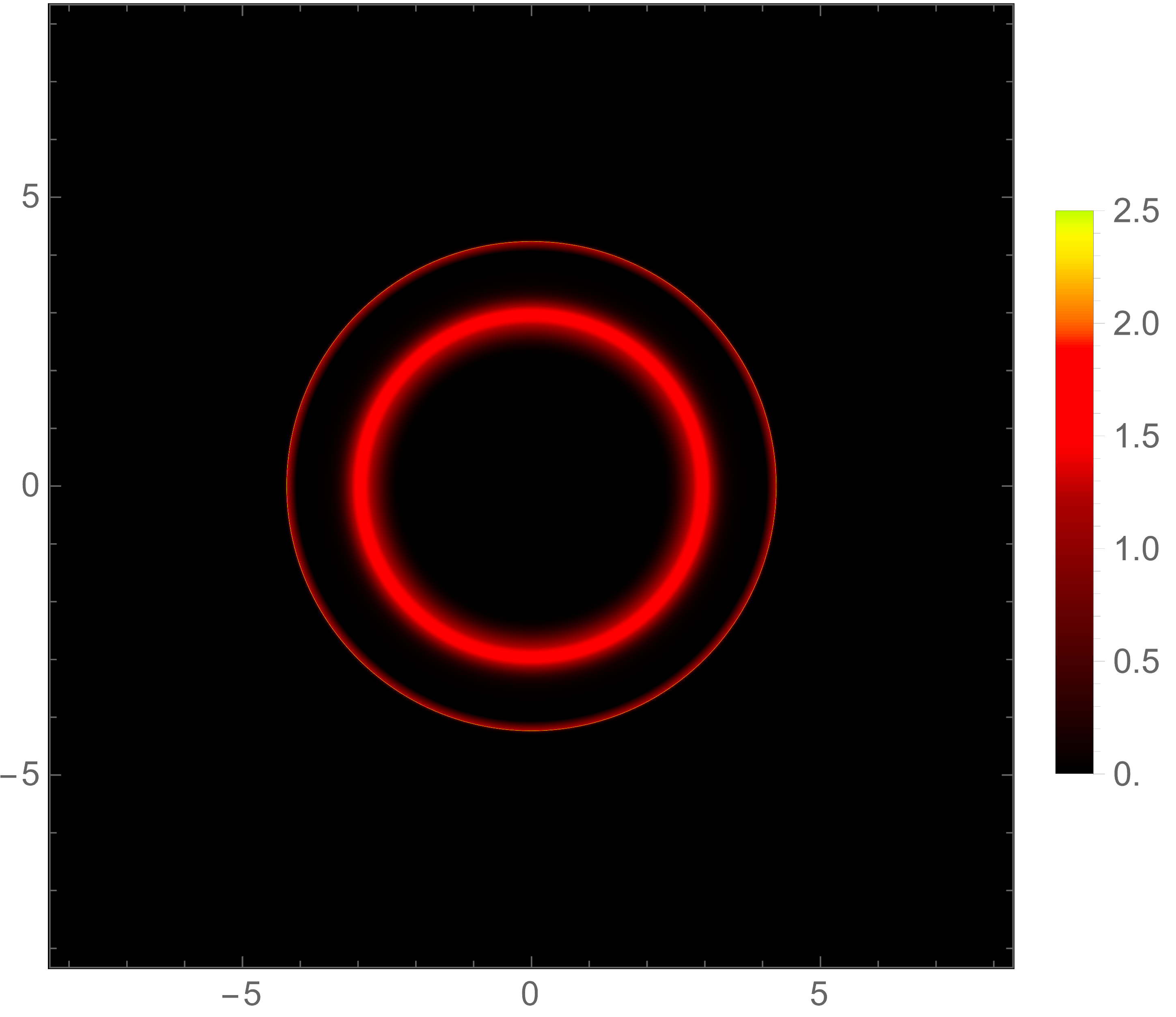}
\includegraphics[width=4cm,height=3cm]{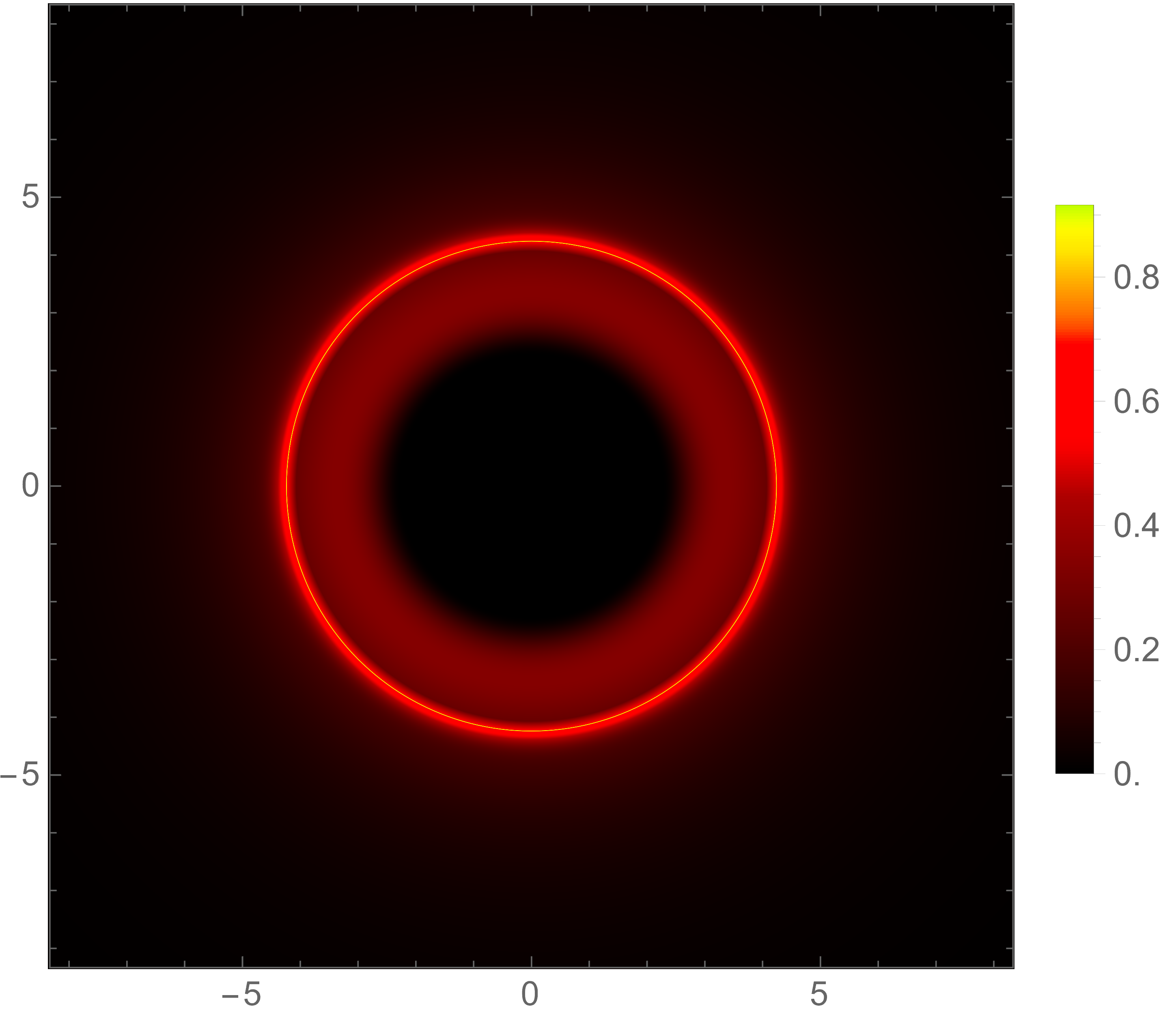}
\includegraphics[width=4cm,height=3cm]{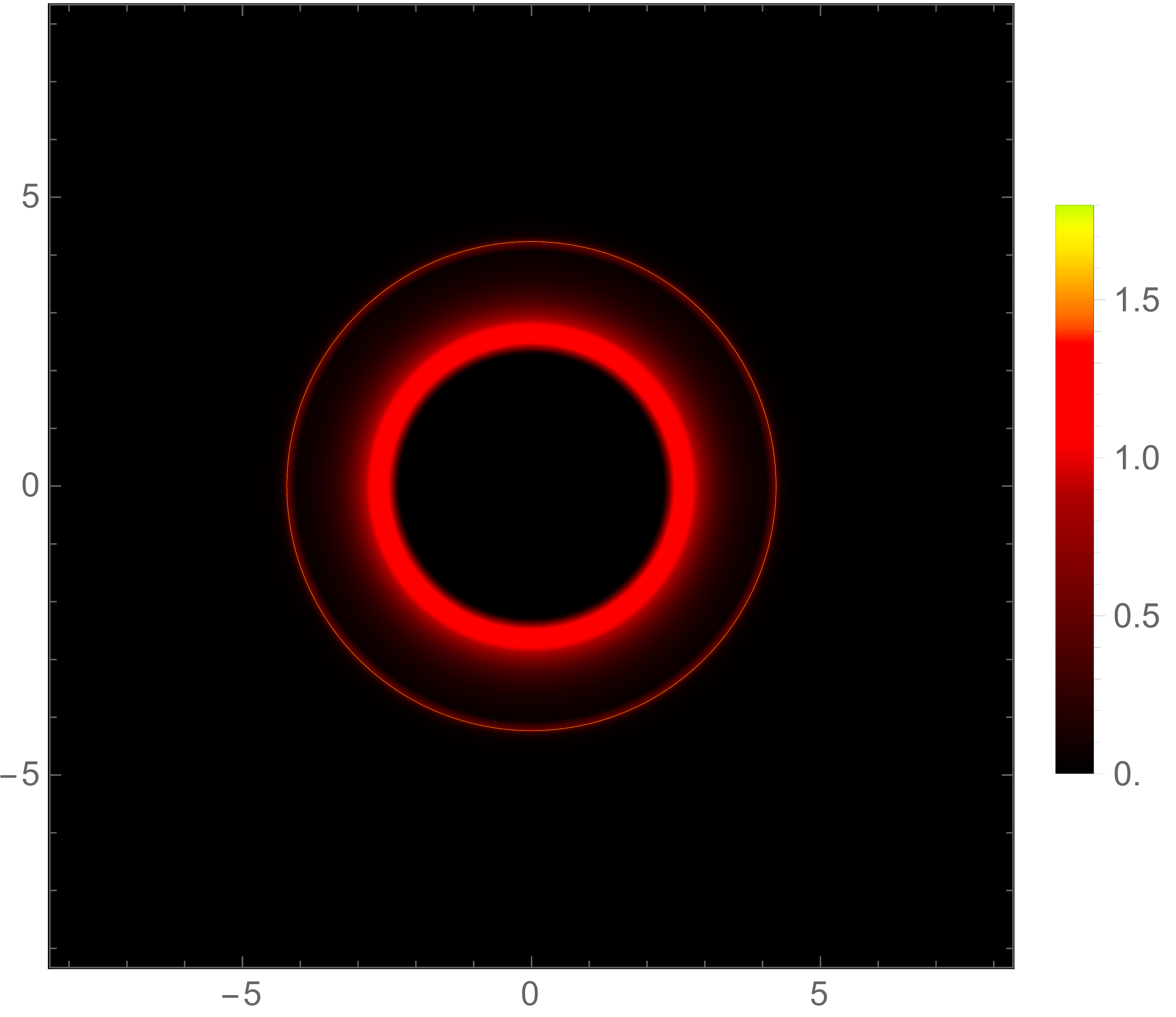}
\includegraphics[width=4cm,height=3cm]{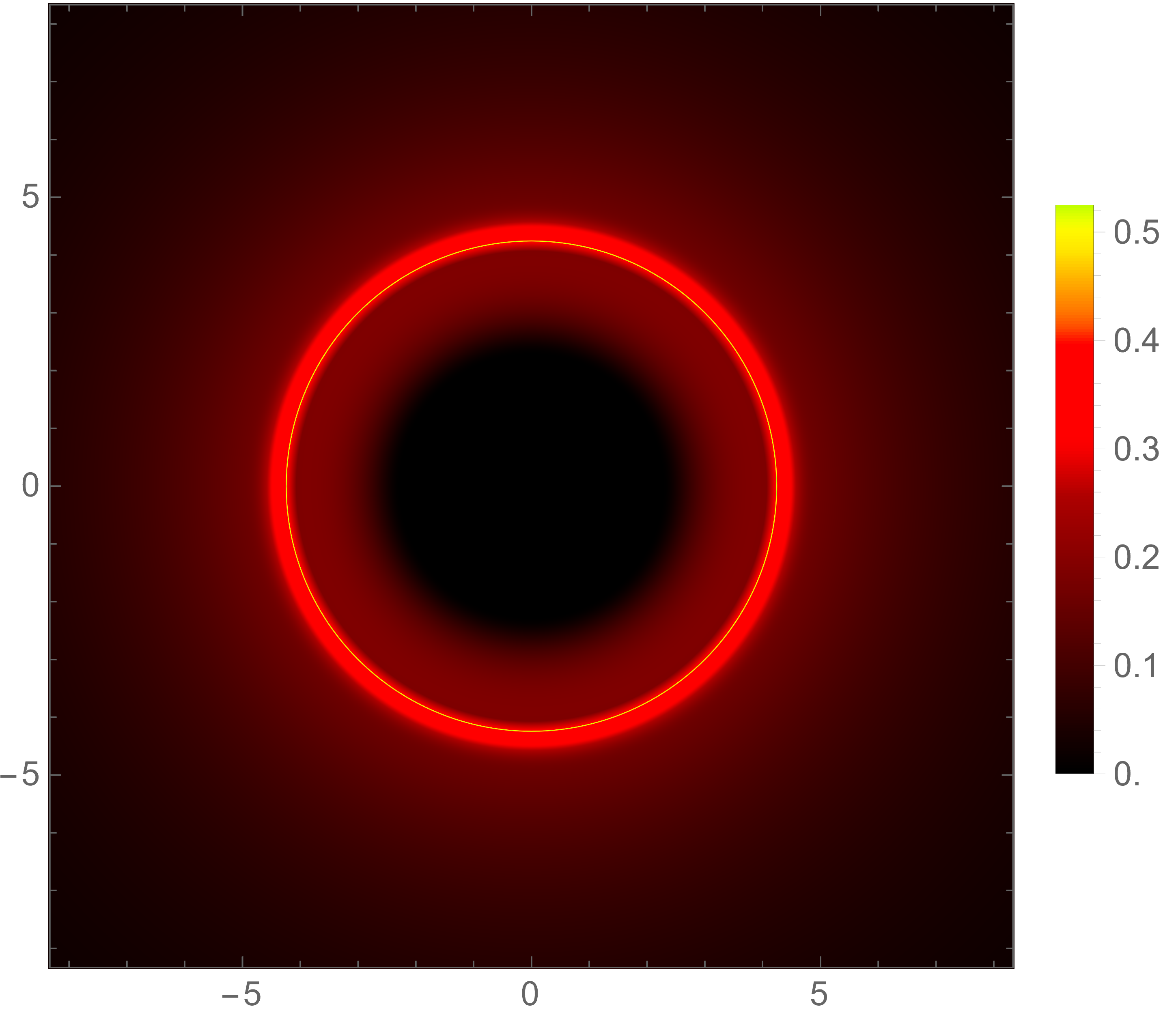}
\includegraphics[width=4cm,height=3cm]{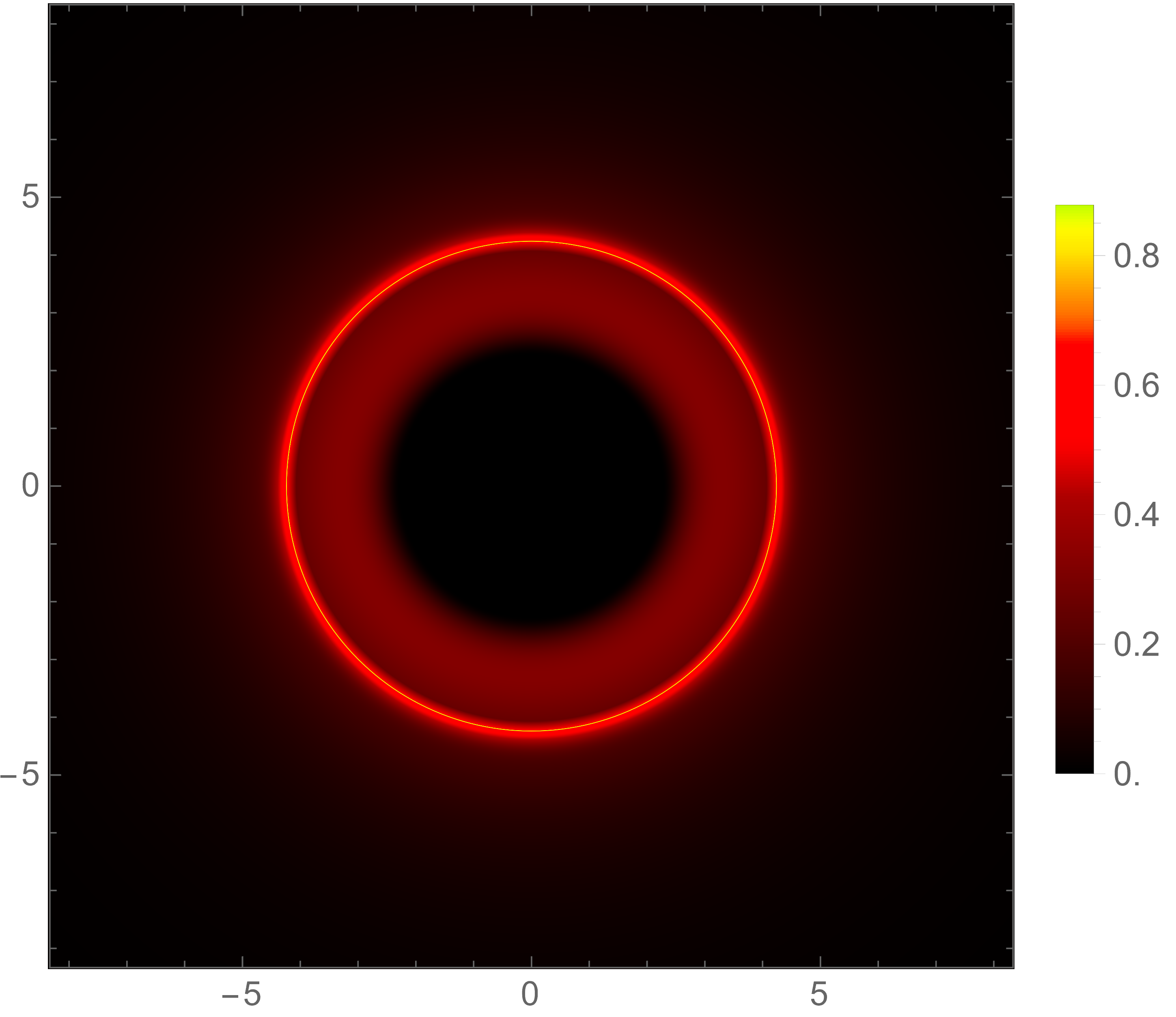}
\includegraphics[width=4cm,height=3cm]{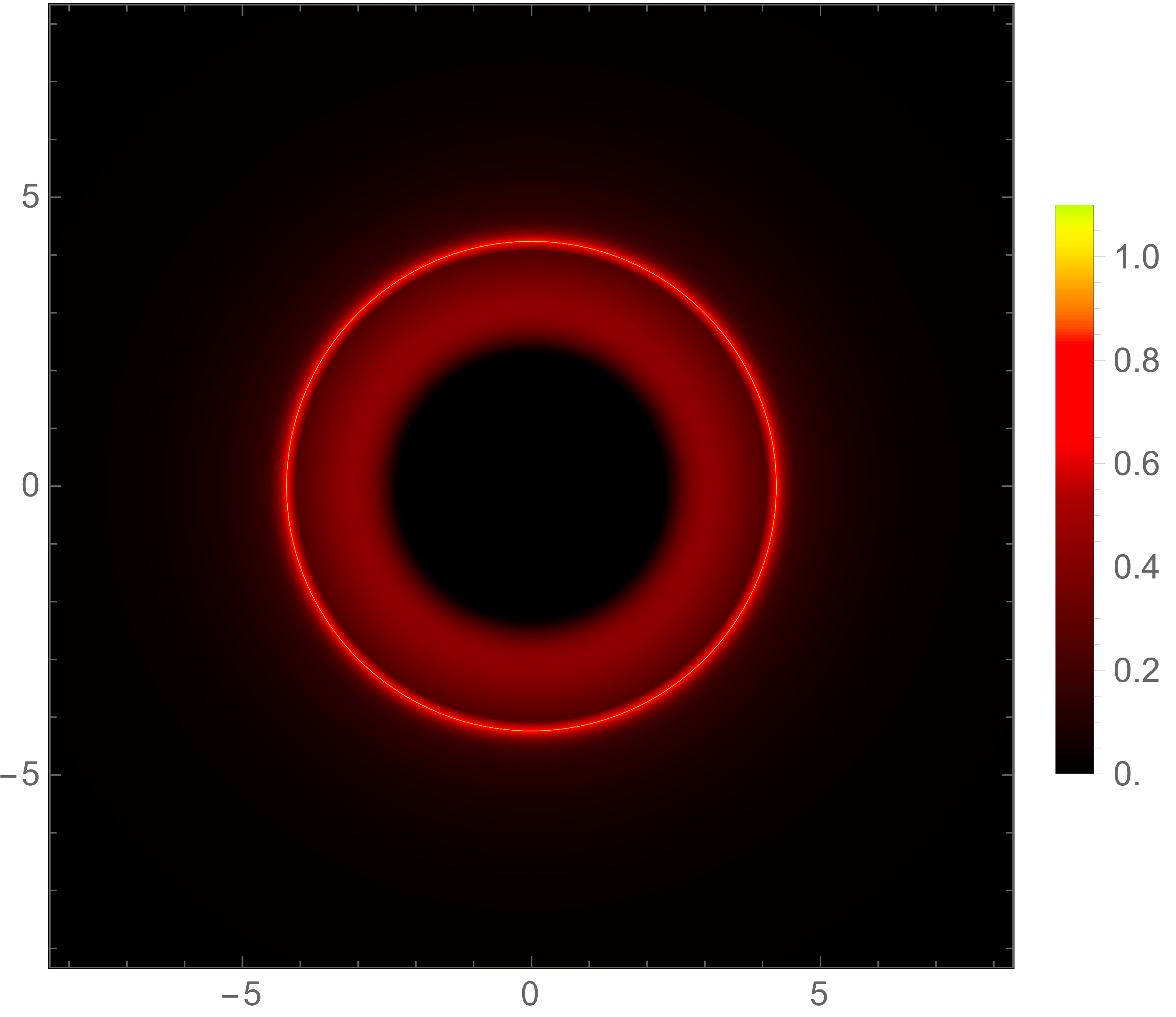}
\includegraphics[width=4cm,height=3cm]{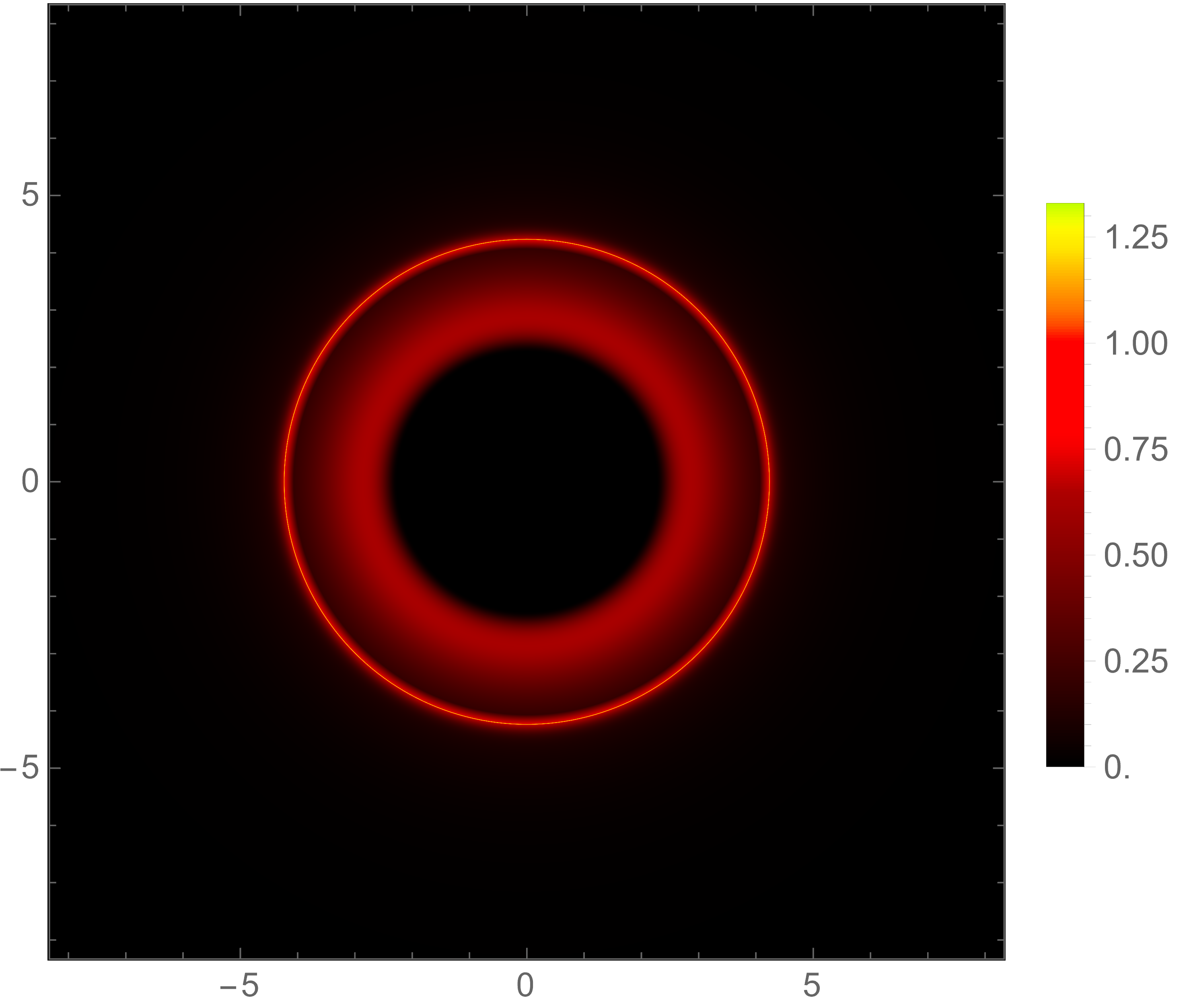}
\caption{Imaging LQG black holes with $P=0.08$ using the ten SU profiles (from left to right and top to bottom) depicted in Fig. \ref{fig:Pro}. For each image, in the vertical bar we depict the relative intensities.}
\label{fig:ima}
\end{figure*}

In Fig. \ref{fig:ima} we depict the result of the imaging of LQG black holes with $P=0.08$ according to the ten SU profiles depicted in Fig. \ref{fig:Pro}. Through this sequence of images we can see the combined effect of the location of the peak of the emission and the strength of the decay with distance. In all cases, the photon ring $n=1$ produces a neat boost of luminosity in the images. However, while in models SU2, SU4, and SU6, such a photon ring appears clearly separated from the direct emission, in the SU4 and SU6 models it lies above the direct $n=0$ emission, and in the SU2 model it appears inside it. This is a reflection of the shape of the emission profile as compared to the location of the photon sphere $r_m$. Similarly, how spread the emission is correlates with how far the luminosity in the image extends to: for instance in SU1 and SU3 models, the disk provides a wide luminosity surrounding the black hole. Finally, the size of the shadow is reduced in those models in which the emission is weak outside the event horizon, as is clearly seen in SU7/SU8/SU9/SU10 models. Indeed, in such models the optical appearance of LQG black holes is very similar, and thus hardly distinguishable from each another.

We point out that such images are quite similar to what one finds in Schwarzschild black holes at equal emission model\footnote{Note, however, that a comparison between both models on complete equal-footing is not possible, given the fact that the locations of their event horizons are not exactly the same, implying that there will be certain differences between the different emissions models. }. One possibility to distinguish between them is via the rate of luminosity between the $n=1$ and $n=2$ rings, since both rings are targets of future projects based in very long baseline interferometry \cite{Johnson:2019ljv}. Such a rate is a universal property of a given background geometry in the limit $n \to \infty$ given the fact that the ratio of luminosity follows the same rule (\ref{lyapunovresuelto}) as for the ratio of locations and it is thus governed by the Lyapunov exponent. However, in real scenarios of the accretion flow photons with different impact parameters that traverse through different regions of disk will be emitted with different luminosities, entailing certain differences with respect to the prediction of the Lyapunov exponent. In our case we compute the ratio of luminosity between the $n=1$ and $n=2$ rings since, given the exponential suppression of luminosity, higher-order rings will contribute negligibly to the total luminosity and, therefore, such a ratio will closely approach the one of the Lyapunov exponent.

We report our findings in Table \ref{tab:rate} for LQG black holes with $P=0.08$ as compared to the values for the Schwarzschild black hole, using the ten SU models. We observe that, in agreement with our findings regarding gravitational lensing of individual light rays, and the smaller Lyapunov index, the luminosity ratio is smaller in LQG black holes than in the Schwarzschild black hole at equal emission SU model. This entails that, similarly to what has been observed in other gravitational geometries in the literature, the luminosity ratio could potentially (and hopefully) allow to distinguish between different black hole geometries should we were able to isolate the contributions of the $n=1$ and $n=2$ photon rings using very long baseline interferometry techniques \cite{Carballo-Rubio:2022aed}.

\begin{table*}[t!]
\begin{tabular}{|c|c|c|c|c|c|c|c|c|c|c|c|}
\hline
 Model & $e^{\gamma}$  & $\left(\frac{I_1}{I_2}\right)^{SU1} $  & $\left(\frac{I_1}{I_2}\right)^{SU2} $ & $\left(\frac{I_1}{I_2}\right)^{SU3} $ & $\left(\frac{I_1}{I_2}\right)^{SU4} $ & $\left(\frac{I_1}{I_2}\right)^{SU5} $ & $\left(\frac{I_1}{I_2}\right)^{SU6} $ & $\left(\frac{I_1}{I_2}\right)^{SU7} $ & $\left(\frac{I_1}{I_2}\right)^{SU8} $ & $\left(\frac{I_1}{I_2}\right)^{SU9} $ & $\left(\frac{I_1}{I_2}\right)^{SU10} $ \\ \hline
Schwarzschild & 23.35 & 28.94 & 27.83 & 21.57 & 27.21 & 23.34  & 21.14 & 24.74 & 23.45 & 22.81 & 22.20 \\ \hline
LQG ($P=0.04$) & 22.07 & 27.75 & 26.65 & 20.33 & 25.96   & 22.17  & 19.97 & 23.48 & 22.19 & 21.56 &  20.95 \\ \hline
LQG ($P=0.08$) & 20.80 & 26.54 & 25.45 & 19.11 & 24.70   & 21.00  & 18.81 & 22.22 & 20.94 & 20.33 &  19.71 \\ \hline
\end{tabular}
\caption{The ratio of luminosities $I_1/I_2$ between the first and second photon rings for LQG black holes with $P=0.04$ and $P=0.08$ as compared  to the Schwarzschild one, for the ten SU models considered in this work. We also depict  the theoretical rate of intensities between photon rings in the $n \to \infty$ limit associated to the Lyapunov exponent on each case, which deviates from the theoretical ratio of luminosities (i.e. for a fully homogeneous disk's emission) of the $n=1$ and $n=2$ photon rings just by a mild $\lesssim 0.3\%$.}
\label{tab:rate}
\end{table*}

\section{Conclusion and discussion} \label{C:VI}

In this work we have considered the gravitational lensing and shadow images of a family of asymptotically-flat, spherically symmetric configurations derived from Loop Quantum Gravity, and characterized by two new scales $a_0$ and $P$. Assuming $a_0$ to be vanishing in order to deal with ordinary black hole solutions, and $P$ to be small enough in order to be consistent with the inferred size of the shadow of the supermassive central object at the heart of the Milky Way galaxy which restrict it to the range $0<P \lesssim 0.08$, we have studied gravitational lensing in both the weak and strong gravitational limits, and used that knowledge to characterize the cast images of the modified LQG black hole by a thin accretion disk. 

In the weak gravitational lensing regime we have found that corrections induced by the LQG black holes slightly the deflection angle within the bounds above. In the strong gravitational lensing regime we have found the corrections to the photon sphere (the locus of unstable bound orbits) and its associated critical impact parameter as well as to the lensing deflection coefficients. Our results indicate that these LQG black holes yield in this regime a stronger growth of the deflection angle as the corresponding impact parameter is approached. Associated to these results, we have computed several observables, including the Lyapunov exponent capturing the radial growth of nearly-bound orbits as well as the angular separation and flux ration of the multiple images. 

Finally we tackled the analysis of the cast images of LQG black holes when illuminated by a geometrically and optically thin accretion disk, modelled by a bunch of semi-analytic profiles for a monochromatic emission in the disk's frame. Besides finding the full images for each configuration, we characterized the ratio of luminosity between the $n=1$ and $n=2$ photon rings, given the fact that they are potential targets of future very long baseline interferometric projects, finding a neat reduction of such a rate for LQG black holes as compared to its Schwarzschild counterpart. This means that, at equal emission model, both models provide sufficiently different signatures to be potentially distinguished.

The bottom line of our analysis is that differences between LQG and Schwarzschild black holes using gravitational lensing, at least within current constraints for the parameter $P$, are potentially within measurable range by future observational devices. Obviously, as technology progresses, more stronger bounds on the parameter $P$ (for instance via the upgrade of resolutions in EHT observations mentioned in this paper) are yet to be expected. On the other hand, here we did not consider configurations with non-vanishing $a_0$, since this constant is expected to be very small within the LQG theory it is derived from, and the wormhole structure it holds would thus be covered by an event horizon. 

\section{Acknowledgments}

This work is supported by the Spanish National Grant PID2022-138607NBI00, funded by MCIN/AEI/10.13039/501100011033 (``ERDF
A way of making Europe" and ``PGC Generaci\'on de
Conocimiento"). This article is based upon work from
COST Action CA21136, funded by COST (European Cooperation in Science and Technology).

\end{document}